\documentclass[preprint,preprintnumbers,amsmath,amssymb]{revtex4}
\usepackage{amssymb}
\usepackage{latexsym}
\usepackage{epsfig}
\begin{document}
\title{Implications, consequences and interpretations of generalized entropy in the cosmological setups}
\author{H. Moradpour\footnote{h.moradpour@riaam.ac.ir}}
\address{Research Institute for Astronomy and Astrophysics of Maragha
(RIAAM), P.O. Box 55134-441, Maragha, Iran}

\begin{abstract}
Recently, it was argued~(Eur. Phys. J. C {\bf73}, 2487 (2013)) that
the total entropy of a gravitational system should be related to the
volume of system instead of the system surface. Here, we show that
this new proposal cannot satisfy the unified first law of
thermodynamics and the Friedmans equation simultaneously, unless the
effects of dark energy candidate on the horizon entropy are
considered. In fact, our study shows that some types of dark energy
candidate may admit this proposal. Some general properties of
required dark energy are also addressed. Moreover, our investigation
shows that this new proposal for entropy, while combined with the
second law of thermodynamics (as the backbone of Verlinde's
proposal), helps us in providing a thermodynamic interpretation for
the difference between the surface and bulk degrees of freedom
which, according to Padmanabhan's proposal, leads to the emergence
of spacetime and thus the universe expansion. In fact, our
investigation shows that the entropy changes of system may be equal
to the difference between the surface and bulk degrees of freedom
falling from surface into the system volume. Briefly, our results
signal us that this new proposal for entropy may be in agreement
with the thermodynamics laws, the Friedmann equation, Padmanabhan's
holographic proposal for the emergence of spacetime and therefore
the universe expansion. In fact, this new definition of entropy may
be used to make a bridge between Verlinde's and Padmanabhan's
proposals.
\end{abstract}

\maketitle
\section{Introduction}
The Bekenstein entropy is the backbone of study the thermodynamic
aspects of Einstein field equations \cite{pois,T11}. Indeed, thanks
to Jacobson unique work \cite{J1}, we can take into account the
Einstein field equations as a thermodynamical equation of state. His
approach also provides a suitable framework for finding out the
thermodynamic equation of state corresponding to the gravitational
field equations in other theories of gravity \cite{J11}. Since the
observed universe is dynamics \cite{roos}, described by
Friedmann-Lemaitre-Robertson-Walker (FLRW) metric, the investigation
of mutual relation between thermodynamics laws and Friedman
equation, governing the universe evolution, is important. In fact,
such generalizations have been addressed by many authors in vast
cosmological setups
\cite{Hay2,Hay22,Bak,CaiKim,Cai2,Caimore1,sheyw1,sheyw2,m}.

There are two definitions for the temperature of apparent horizon of
FLRW universe \cite{Hay2,Hay22,Bak,CaiKim,CaiKimt,GSL1}, called the
Hayward-Kodama temperature \cite{Hay2,Hay22,Bak} and the Cai-Kim
temperature \cite{CaiKim,CaiKimt}, and both of them, whenever they
are combined with the Bekenstein entropy relation, are in agreement
with the Friedmann equation as well as the unified first law of
thermodynamics.  It is useful to note here that although the
Bekenstein entropy is in line with the current accelerating phase of
universe expansion and the second and generalized second laws of
thermodynamics \cite{msrw,mr}, some authors show that a dark energy
candidate, due to its unknown nature, may modify the horizon entropy
(the Bekenstein limit in the Einstein general relativity framework)
\cite{cana,cana1,em,md,mm}. In their setups, the Friedmann equation
and the second law of thermodynamics are still valid simultaneously
\cite{cana,cana1,em,md,mm}. Finally, it is useful to mention here
that one may get the universe evolution equation in various theories
of gravity (the corresponding Friedmann equation), by taking into
account the horizon entropy relation together with the
Hayward-Kodama temperature and applying the unified first law of
thermodynamics on the apparent horizon of FLRW universe \cite{AA}.

Verlinde shows that the second law of thermodynamics (tendency of
systems to increase their entropy \cite{CALLEN}) may lead to the
emergence of spacetime and gravity \cite{Ver}. The entropy-area
relation (the Bekenstein entropy for General relativity) plays a key
role in this theory which attracts more attempts to itself
\cite{Cai4,Cai41,Smolin,Li,Tian,Myung1,Vancea,Modesto,Sheykhi1,BLi,Sheykhi2,Sheykhi21,Sheykhi22,Sheykhi23,Ling,Sheykhi24,Gu,Miao1,other,mann,SMR,ms}.
In another approach, Padmanabhan argues that the difference between
the surface and bulk degrees of freedom is proportional to the
volume changes of system leading to the emergence of spacetime and
thus the universe expansion \cite{pad1}. Indeed, the positive
difference between the surface and bulk degrees of freedom yields
the spacetime emergence and the Friedmann equations in various
theories of gravity \cite{pad1,pad10,pad2,pad3,jalal1}. It is also
useful to note here that the Padmanabhan argument claims that the
Cai-Kim temperature is more suitable option for the temperature of
matter fields enclosed by the apparent horizon of FLRW universe
\cite{pad1}. Now, one can ask that why is the difference between the
surface and bulk degrees of freedom positive, leading to the
emergence of spacetime and universe expansion? Moreover, bearing the
Verlinde argument in mind, may one relate the positive difference
between the surface and bulk degrees of freedom to the second law of
thermodynamics? Briefly, what is the relation between the
thermodynamic laws and Padmanabhan's proposal?

Recently, Tsallis and Cirto proposed a new expression for the
horizon entropy of Schwarzschild black hole, and therefore the
entropy-area relation \cite{salis}. Indeed, they have considered the
general formalism of non-additive entropy, applied that to the
Schwarzschild horizon and obtained that the horizon entropy is
related to the volume confined by it, instead of its surface area
\cite{salis}. Moreover, It was shown that if one uses the
Weyl-Wigner-Groenewold-Moyal formalism of deformation quantization,
then the apparent horizon entropy of radiation and dust dominated
quantum universes is proportional to the universe volume instead of
its surface area \cite{jalal}. Therefore, both the non-extensive
statistics and quantum cosmology theories suggest that the horizon
entropy is related to the system volume instead of its surface area.
Therefore, since it was shown that the Einstein field equations are
in line with thermodynamics if the horizon entropy be proportional
with its surface
\cite{T11,J1,J11,Hay2,Hay22,Bak,CaiKim,CaiKimt,GSL1}, it seems that
this new hypothesis for the entropy-area relation, given by both
quantum cosmology considerations \cite{jalal} and non-extensive
statistics \cite{salis}, leads to an inconsistency in the mutual
relation between the Einstein field equations and the thermodynamics
laws. This inconsistency necessitates us to more investigate the
relation between the non-extensive thermodynamics and the
gravitational field equations. In refs.~\cite{salis1,salis2,salis3},
authors show the Tsallis et al. entropy, called the generalized
entropy, may be related to the bulk viscosity of cosmological fluids
leading to modify the Friedmann equation. It is also pointed out
that the entropy of holographic screen with entangled bits meets the
generalized entropy relation and may confirm Verlinde's hypothesis
of gravity \cite{salis4}. Moreover, it seems that the generalized
entropy, whenever it is combined with Verlind's formalism, modifies
gravitational coupling constant and thus the acceleration formula
\cite{salis5,s}. More studies on the relation between the
generalized entropy and various aspects of cosmos can be found in
\cite{salis6,salis7,salis8}.

Therefore, based on some previous works, the Bekenstein entropy (as
the geometrical entropy of horizon) is equal to the maximum entropy
of fields confined by the system boundary \cite{pois,T11}. Now,
since there are some works showing that a dark energy candidate may
modify the horizon entropy \cite{cana,cana1,em,md,mm}, can one make
a relation between the energy density of dark energy candidate and
the general entropy relation? Here, we are going to investigate the
possibility of establishing a relation between these attempts and
the generalized entropy proposed by Tsallis et al.~\cite{salis},
which is also supported by the quantum cosmology considerations
\cite{jalal}. The latter may lead us to find out a profile density
for the dark energy candidate. Moreover, we are eager to study the
relation between the second law of thermodynamics (tendency of
systems to increase their entropy) as the backbone of Verlinde's
proposal, the generalized entropy and Padmanabhan's hypothesis about
the origin of spacetime and the universe expansion. In other words,
we try to show that the generalized entropy, combined by the second
law of thermodynamics, is in line with Padmanabhan's proposal. In
addition, we show that the combination of the second law of
thermodynamics and the generalized entropy may lead to an acceptable
thermodynamic interpretation for the positive difference between the
surface and bulk degrees of freedom.

In order to present our work, we organize the paper as follows.
Bearing the Cai-Kim temperature in mind, we apply the unified first
law of thermodynamics to the apparent horizon of FLRW metric, and
get a relation between the horizon entropy and the energy density of
fluid enclosed by horizon, in the next section. We also point out
the results of considering the generalized and Bekenstein entropies
in the obtained relation. In addition, by making a connection
between the generalized entropy, the Bekenstein entropy and the
energy density of the dark energy candidate, we get a relation for
the energy density of dark energy candidate. Finally, we also study
some general behavior of the obtained dark energy candidate. In
section ($\textmd{III}$), we investigate the relation between the
generalized entropy, the second law of thermodynamics (as the
backbone of Verlinde's proposal) and Padmanabhan's proposal. We show
that the generalized entropy may be used to build a bridge between
the second law of thermodynamics and the positive difference between
the surface and bulk degrees of freedom (as the key point of
Padmanabhan's proposal). The last section is devoted to a summary
and concluding remarks. Throughout this paper we also set
$G=\hbar=c=k_B=1$ for the sake of simplicity.

\section{Thermodynamics, Friedmann equation, the generalized entropy and dark energy}
Consider a FLRW metric
\begin{eqnarray}\label{frw}
ds^{2}=-dt^{2}+a^{2}\left( t\right) \left[ \frac{dr^{2}}{1-\kappa r^{2}}%
+r^{2}d\Omega^{2}\right],
\end{eqnarray}
in which $a(t)$ is scale factor, and $\kappa=-1,0,1$, called the
curvature constant, points to the open, flat and closed universes,
respectively \cite{roos}. The radii of marginally trapped surface,
called the apparent horizon, is defined as
\begin{eqnarray}\label{ah2}
\partial_{\alpha}\zeta\partial^{\alpha}\zeta=0\rightarrow r_A,
\end{eqnarray}
where $\zeta=a(t)r$, and by some calculations one gets
\begin{eqnarray}\label{ah}
\tilde{r}_A=a(t)r_A=\frac{1}{\sqrt{H^2+\frac{\kappa}{a(t)^2}}},
\end{eqnarray}
for the physical radii of apparent horizon ($\tilde{r}_A$)
\cite{Hay2,Hay22,Bak,sheyw1,sheyw2}. If this spacetime be filled by
a prefect fluid with energy-momentum tensor $T^{\nu}_{\
\mu}=\textmd{diag}(-\rho,p,p,p)$, the corresponding continuity
equation can be written as
\begin{equation}\label{cont}
\dot{\rho}+3H(\rho+p)=0,
\end{equation}
where $\rho$ and $p$ are the energy density and pressure of the
assumed source. $H\equiv\frac{\dot{a}}{a}$ is also called the Hubble
parameter. Since it is argued that the apparent horizon can be
considered as a proper causal boundary for the FLRW spacetime
\cite{Hay2,Hay22,Bak,sheyw1,sheyw2}, we take into account it as the
causal boundary. The energy amount crossing the apparent horizon
during the universe expansion ($\delta Q^m$) is defined as
\begin{eqnarray}\label{uf1}
\delta Q^m=A(T^b_a\partial_b \zeta + W\partial_a \zeta)dx^a,
\end{eqnarray}
where $W=\frac{\rho-p}{2}$ and $\zeta=\tilde{r}_A$ are the work
density and apparent horizon radii, respectively \cite{Cai2}. After
some calculations and using Eq.~(\ref{cont}), one obtains
\cite{md,mm}
\begin{eqnarray}\label{ufl2}
\delta Q^m=-3VH(\rho+p)dt=Vd\rho.
\end{eqnarray}
In deriving this equation, we adopt $V=\frac{4\pi}{3}\tilde{r}_A^3$
together with the Cai-Kim approach, in which $d\zeta \approx 0$, in
the infinitesimal time $dt$, \cite{CaiKim}. Bearing the Clausius
relation in mind ($TdS_A=\delta Q^m$) \cite{CALLEN}, we get
\begin{eqnarray}\label{ufl3}
d\rho=-\frac{T}{V}dS_A,
\end{eqnarray}
in that $S_A$ is the horizon entropy, and mines sign comes from the
universe expansion \cite{CaiKim,Cai2,em,md,mm}. Now, for a theory in
which $S_A=S_A(A)$, since the Cai-Kim temperature meets the
$T=\frac{1}{2\pi\tilde{r}_A}$ relation
\cite{CaiKim,Cai2,CaiKimt,GSL1}, one reaches
\begin{eqnarray}\label{ufl4}
\rho=-6\int \frac{S_A^{\prime}(A)}{A^2}dA,
\end{eqnarray}
where $^{\prime}$ denotes derivative with respect to $A$. It is
useful to note here that this result is also obtained if one uses
the Hayward-Kodama temperature \cite{AA}. If we insert Bekenstein
relation ($S_B=\frac{A}{4}$), as the apparent horizon entropy, into
this equation and take integral from the RHS, we get
\begin{equation}\label{freid2}
\frac{1}{\tilde{r}_A^2}=\frac{8\pi}{3}\rho,
\end{equation}
which can finally be rewritten as
\begin{equation}\label{freid1}
H^2+\frac{\kappa}{a^2}=\frac{8\pi}{3}\rho,
\end{equation}
by helping Eq.~(\ref{ah}). Indeed, it is nothing but the Friedmann
equation \cite{roos,CaiKim,Cai2,Caimore1}. Briefly, one can use the
unified first law of thermodynamics~(\ref{uf1}), the Cai-Kim
approach, the Clausius relation and the Bekenstein entropy relation
(as a candidate for the total entropy of system) to get an
equation~(\ref{ufl4}) which may govern the Friedmann
equation~(\ref{freid1}) \cite{CaiKim,Cai2,Caimore1}.

Recently, Tsallis and et al. used the generalized nonadditive
entropy to get a new relation for the Schwarzschild black hole
entropy \cite{salis}. Based on their proposal, the entropy of a
$3+1$-dimensional black hole ($S_A$) is
\begin{eqnarray}\label{ents}
S_A=\gamma A^{\frac{3}{2}},
\end{eqnarray}
where $\gamma$ is the proportionality constant, and should be
evaluated from the dimensional analysis and other parts of physics
\cite{salis}. It is worth mentioning that because entropy is a
non-zero positive quantity \cite{CALLEN}, $\gamma$ should be a
non-zero positive quantity. Since the apparent horizon is a
holographic surface and it may play the role of boundary for
cosmological setup \cite{Hay2,Hay22,Bak}, we generalize this entropy
relation to the apparent horizon of FLRW metric and study its
results. Indeed, since de-Sitter spacetime includes a static black
hole, it is a static spacetime, such generalization for the
de-Sitter spacetime is not far from reality. Moreover, it was shown
that the apparent horizon entropy in radiation and dust dominated
quantum universes also meets Eq.~(\ref{ents}) \cite{jalal}.
Therefore, our generalization is not far-fetched. Now, by inserting
this relation into Eq.~(\ref{ufl4}), and using Eq.~(\ref{ah}), we
reach at
\begin{eqnarray}\label{friedsal1}
\rho=\frac{9\gamma}{\sqrt{\pi}}\sqrt{H^2+\frac{\kappa}{a^2}}.
\end{eqnarray}
Because this equation differs from the Friedmann
equation~(\ref{freid1}), we conclude that if we accept the Friedmann
equation and relation~(\ref{ents}) for the apparent horizon entropy,
then relation~(\ref{ents}) does not lead to satisfaction of the
unified first law of thermodynamics. Therefore, one can blame all of
the assumptions and recipe leading to Eq.~(\ref{ufl4}), and also
similar attempts such as those introduced in
Refs.~\cite{CaiKim,Cai2,GSL1,AA}. Briefly, it seems that there is an
inconsistency between the generalized entropy relation, Friedmann
equation and the unified first law of thermodynamics.

It was also argued that a dark energy candidate, due to its unknown
nature, may modify the apparent horizon entropy in various theories
of gravity \cite{cana,cana1,em,md,mm}. In fact, by decomposing
$\rho$ to $\rho_D+\rho_o$, in which $\rho_D$ denotes the energy
density of dark energy candidate and $\rho_o$ includes the energy
density of other parts of cosmos, using the Friedman
equation~(\ref{freid1}), continuity equation~(\ref{cont}) together
with the unified first law of thermodynamics and the Clausius
relation, one can obtain a new relation for the horizon entropy as
\begin{eqnarray}\label{general1}
S_A=S_B+\frac{1}{6}\int A^2 d\rho_D,
\end{eqnarray}
while dark energy does not interact with other parts of cosmos
\cite{em,md}. Now, by combining this equation with Eq.~(\ref{ents}),
and after some calculation, we get
\begin{eqnarray}\label{dark1}
\rho_D=\frac{3H^2}{8\pi}-\frac{9\gamma}{\sqrt{\pi}}H,
\end{eqnarray}
for the energy density of dark energy candidate in a flat FLRW
universe ($\kappa=0$). It is useful to note here that there are
various models for dark energy in which the density of dark energy
candidate is similar to this result
\cite{ven,ven1,ven2,saha,sola1,sola2,sola3,ven3,sola,GGDE1,sola4,ven4,GGDE,lima}.
Moreover, for a FLRW universe with arbitrary curvature parameter
$\kappa$, we reach
\begin{eqnarray}\label{dark2}
\rho_D=\frac{3(H^2+\frac{\kappa}{a^2})}{8\pi}-\frac{9\gamma}{\sqrt{\pi}}\sqrt{H^2+\frac{\kappa}{a^2}},
\end{eqnarray}
which is also similar to some previous works
\cite{cana,shapiro,sol,gran,dgp}, and converges to Eq.~(\ref{dark1})
in the appropriate limit ($\kappa=0$). Therefore, a dark energy
candidate with profile density, satisfying Eq.~(\ref{dark2}), may be
considered as an origin for the generalized entropy relation.
\subsection*{Some properties of the obtained dark energy candidate}

Considering $V=\frac{4\pi}{3}\tilde{r}_A^3$, by combining
Eqs.~(\ref{ah2}) and~(\ref{dark2}) with each other, we get
\begin{eqnarray}
\rho_D=\alpha V^{-\frac{2}{3}}+\beta V^{-\frac{1}{3}},
\end{eqnarray}
in which $\alpha=(\frac{4\pi}{3})^{\frac{2}{3}}\frac{3}{8\pi}$ and
$\beta=-(\frac{4\pi}{3})^{\frac{2}{3}}\frac{9\gamma}{\sqrt{\pi}}$.
Bearing the $p_D=-\frac{\partial E_D}{\partial V}$ relation in mind,
where $p_D$ and $E_D=\rho_DV$ denote the pressure and energy of dark
energy candidate, respectively, simple calculation leads to
\begin{eqnarray}
p_D=-\frac{1}{3}(\rho_D+\beta V^{-\frac{1}{3}}),
\end{eqnarray}
for the dark energy candidate pressure. For the state parameter
$w_D=\frac{p_D}{\rho_D}$, one also reaches
\begin{eqnarray}\label{state}
w_D=-\frac{1}{3}(1+\frac{\beta V^{\frac{1}{3}}}{\alpha+\beta
V^{\frac{1}{3}}}).
\end{eqnarray}
It was shown that in order to describe the current acceleration
phase of universe expansion ($\ddot{a}\geq0$), the dark energy state
parameter should meet the $w_D\leq-\frac{1}{3}$ condition \cite{mr}.
Applying this condition to Eq.~(\ref{state}), we get the
$\gamma\geq\gamma_0=\frac{1}{24\tilde{r}_A\sqrt{\pi}}(\frac{3}{4\pi})^\frac{1}{3}$
condition for the gamma parameter. Therefore, the minimum value of
$\tilde{r}_A$ may be used to choose a suitable value for $\gamma$.
As an example, for a flat universe of $w_D=-1$, it is easy to show
that
$\gamma_{[w_D=-1]}=\frac{H}{12\sqrt{\pi}}(\frac{3}{4\pi})^\frac{1}{3}$.
Since for a flat universe $\tilde{r}_A=\frac{1}{H}$, we can see
that, as a desired result, $\gamma_{[w_D=-1]}>\gamma_0$.
\section{Emergence of spacetime and the universe expansion}

According to Verlinde's hypothesis, the tendency of systems to
increase their entropy (the second law of thermodynamics) leads to
the emergence of spacetime and gravity \cite{Ver}. In this approach,
the Bekenstein entropy and the energy definition play the key role
in getting the gravitational field equations
\cite{Cai4,Cai41,Smolin,Li,Tian,Myung1,Vancea,Modesto,Sheykhi1,BLi,Sheykhi2,Sheykhi21,Sheykhi22,Sheykhi23,Ling,Sheykhi24,Gu,Miao1,other,mann,SMR,ms}.
In addition, Padmanabhan argues that the positive difference between
the surface and bulk degrees freedom may lead to emergence of
spacetime and thus the universe expansion
\cite{pad1,pad10,pad2,pad3,jalal1}. Although the surface degrees of
freedom are proportional with the Bekenstein entropy, the role of
the second law of thermodynamics is not clear in Padmanabhan's
approach. In fact, based on Padmanabhan's argument
$\frac{dV}{dt}\propto(N_{s}-N_{b})$ in which $N_s=A=4S_B$ and
$N_b=\frac{2|E|}{T}$ denote the surface and bulk degrees of freedom,
respectively \cite{pad1}. Moreover, $E=(\rho+3p)V$ is the Komar
mass, while $T$ is the Cai-Kim temperature \cite{CaiKim,CaiKimt}.
$\rho$ and $p$ are also the energy density and pressure of a source
filling the background, respectively
\cite{pad1,pad10,pad2,pad3,jalal1}. Finally, by equating the
proportionality constant to one, Padmanabhan proposes \cite{pad1}
\begin{eqnarray}\label{pad0}
\frac{dV}{dt}=N_{s}-N_{b}.
\end{eqnarray}
Bearing the continuity equation~(\ref{cont}) in mind, it is a matter
of calculation to show that this hypothesis leads to the Friedman
equation in various theories of gravity
\cite{pad1,pad10,pad2,pad3,jalal1}. Now, focus on the generalized
entropy ($S_A=3\gamma\sqrt{4\pi}V$), we may interpret that the
tendency of systems to increase their entropy (the second law of
thermodynamics) implies $\dot{V}\geq0$. Therefore, if we accept
Padmanabhan's proposal ($\frac{dV}{dt}\propto(N_{s}-N_{b})$), we
conclude that the second law of thermodynamics implies $N_s\geq
N_b$. Moreover, if we take the proportionality constant equal to
$3\gamma\sqrt{4\pi}$, we get
\begin{eqnarray}\label{pad}
\dot{S}_A=3\gamma\sqrt{4\pi}\frac{dV}{dt}=3\gamma\sqrt{4\pi}(N_{s}-N_{b}),
\end{eqnarray}
leading to $\frac{dV}{dt}=N_{s}-N_{b}$. Based on Eq.~(\ref{pad}),
the entropy changes of system is proportional with the degrees of
freedom which fall from the system surface into its volume. The
number of these falling degrees of freedom is equal to the
difference between the surface and bulk degrees of freedom. Finally,
we see that the second law of thermodynamics implies a positive
difference between the surface and bulk degrees of freedom, which
leads to the emergence of spacetime and the Friedmann equations in
various theories of gravity \cite{pad1,pad10,pad2,pad3,jalal1}.
Briefly, we saw that the generalized entropy is in agreement with
the second law of thermodynamics (Verlinde's approach) and
Padmanabhan's hypothesis. Indeed, the generalized entropy may help
us in making a bridge between the systems tendency to increase their
entropy (the second law of thermodynamics) as the origin of the
emergence of spacetime in Verlinde's hypothesis and Padmanabhan's
approach.
\section{Summary and concluding remarks}
After referring to the Tsallis et al. proposal for the horizon
entropy of the Schwarzschild black hole, we have generalized it to
the apparent horizon of the FLRW universe and pointed out the
inconsistency between the generalized entropy relation, the
Friedmann equations and the unified first law of thermodynamics. As
we have addressed, our generalization is supported by the quantum
cosmology considerations \cite{jalal}. Additionally, in order to
eliminate this inconsistency, we took into account the works in
which authors claim that a dark energy candidate may modify the
horizon entropy and thus the Bekenstein limit in the Einstein
framework. In continue, we could get an expression for the energy
density of dark energy candidate. Moreover, by considering the
pressure definition in thermodynamics, we got a relation for the
pressure and state parameter of the obtained dark energy candidate.
Our study shows that this dark energy candidate may be used to
describe the current phase of universe expansion. Finally, we showed
that the new proposal for the horizon entropy may be used to make a
bridge between the tendency of systems to increase their entropy
(the second law of thermodynamics) as the backbone of Verlinde's
proposal, and the positive difference between the surface and bulk
degrees of freedom as the origin of the spacetime emergence and its
expansion in Padmanabhan's approach. The latter helps us in getting
more close to the thermodynamic roots of Padmanabhan's approach.
Briefly, our investigation shows that the generalized entropy may
help us in taking into account the second law of thermodynamics as
the cause of positive difference between the surface and bulk
degrees of freedom. In fact, by combining the generalized entropy
relation, the second law of thermodynamics and Padmanabhan's
proposal with each other, we saw that the volume changes of system
is equal to the difference between the surface and bulk degrees of
freedom. The spacetime emerges with the fall of these degrees of
freedom from the system surface into its volume, a mechanism which
may guarantee the universe expansion.
\section*{Acknowledgments}This work has been supported financially by
Research Institute for Astronomy \& Astrophysics of Maragha (RIAAM).

\end{document}